\documentclass[prl,a4paper,twocolumn,showpacs,floatfix,superscriptaddress,amsmath,amsfonts,amssymb,preprintnumbers,citeautoscript]{revtex4}
\pdfoutput=1
\usepackage[T1]{fontenc}\usepackage[latin1]{inputenc}
\usepackage{dcolumn,graphicx,color,booktabs,microtype,afterpage} \graphicspath{{Figures/}}
\usepackage[charter,greekuppercase=italicized]{mathdesign}\usepackage{sidecap}
\renewcommand{\figurename}{Fig.}
\renewcommand{\tablename}{Table}
\makeatletter\renewcommand{\fnum@figure}[1]{\figurename~\thefigure~(color online).}\makeatother
\makeatletter\renewcommand{\fnum@table}[1]{\tablename~\thetable.}\makeatother

\newcount\hh \newcount\mm
\hh=\time \divide\hh by 60
\mm=\hh \multiply\mm by 60 \mm=-\mm
\advance\mm by \time
\def\now{\number\hh:\ifnum\mm<10{}0\fi\number\mm}

\usepackage[colorlinks,plainpages=false,linkcolor=blue,urlcolor=blue,citecolor=blue,pdfpagemode=UseNone,pdfstartview=FitBH]{hyperref}

\newcommand{\half}{\frac{1}{\protect\raisebox{0pt}{\scriptsize 2}}}

\begin{document}

\makeatletter\renewcommand{\ps@plain}{%
\def\@evenhead{\hfill\itshape\rightmark}%
\def\@oddhead{\itshape\leftmark\hfill}%
\renewcommand{\@evenfoot}{\hfill\small{--~\thepage~--}\hfill}%
\renewcommand{\@oddfoot}{\hfill\small{--~\thepage~--}\hfill}%
}\makeatother\pagestyle{plain}


\title{~\vspace*{-12pt}\\Momentum-space structure of quasielastic spin fluctuations in Ce$_3$Pd$_{20}$Si$_6$}

\author{P.~Y.~Portnichenko}
\affiliation{Institut für Festkörperphysik, TU Dresden, D-01069 Dresden, Germany}

\author{A.~S.~Cameron}
\affiliation{Institut für Festkörperphysik, TU Dresden, D-01069 Dresden, Germany}

\author{M.~A.~Surmach}
\affiliation{Institut für Festkörperphysik, TU Dresden, D-01069 Dresden, Germany}

\author{P.~P.~Deen}
\affiliation{European Spallation Source ESS AB, Stora Algatan 4, SE-22100 Lund, Sweden.}
\affiliation{Niels Bohr Institutet, University of Copenhagen, Blegdamsvej 17, DK-2100 Copenhagen, Denmark.}

\author{S.~Paschen}
\affiliation{Institute of Solid State Physics, Vienna University of Technology, Wiedner Hauptstr. 8--10, A-1040 Vienna, Austria}

\author{A.~Prokofiev}
\affiliation{Institute of Solid State Physics, Vienna University of Technology, Wiedner Hauptstr. 8--10, A-1040 Vienna, Austria}

\author{J.-M. Mignot}
\affiliation{Laboratoire L\'{e}on Brillouin, CEA-CNRS, CEA/Saclay, F-91191 Gif sur Yvette, France}

\author{A.~M.~Strydom}
\affiliation{Physics Department, University of Johannesburg, PO Box 524, Auckland Park 2006, South Africa}

\author{M.\,T.~F.~Telling}
\affiliation{ISIS Facility, Rutherford Appleton Laboratory, Chilton, Didcot, Oxon, OX110QX, United Kingdom}
\affiliation{Department of Materials, University of Oxford, Parks Road, Oxford, OX1 3PH, United Kingdom}

\author{A.~Podlesnyak}
\affiliation{Quantum Condensed Matter Division, Oak Ridge National Laboratory, Oak Ridge, TN~37831}

\author{D.~S.~Inosov}\email[Corresponding author: \vspace{5pt}]{Dmytro.Inosov@tu-dresden.de}
\affiliation{Institut für Festkörperphysik, TU Dresden, D-01069 Dresden, Germany}

\begin{abstract}
\noindent Among heavy-fermion metals, Ce$_3$Pd$_{20}$Si$_6$ is one of the heaviest-electron systems known to date. Here we used high-resolution neutron spectroscopy to observe low-energy magnetic scattering from a single crystal of this compound in the paramagnetic state. We investigated its temperature dependence and distribution in momentum space, which was not accessible in earlier measurements on polycrystalline samples. At low temperatures, a quasielastic magnetic response with a half-width $\Gamma\approx0.1$\,meV persists with varying intensity all over the Brillouin zone. It forms a broad hump centered at the $(111)$ scattering vector, surrounded by minima of intensity at $(002)$, $(220)$ and equivalent wave vectors. The momentum-space structure distinguishes this signal from a simple crystal-field excitation at 0.31\,meV, suggested previously, and rather lets us ascribe it to short-range dynamical correlations between the neighboring Ce ions, mediated by the itinerant heavy $f\!$-electrons via the RKKY mechanism. With increasing temperature, the energy width of the signal follows the conventional $T^{1/2}$ law, $\Gamma(T)=\Gamma_0+A\sqrt{T}$. The momentum-space symmetry of the quasielastic response suggests that it stems from the simple-cubic Ce sublattice occupying the 8c Wyckoff site, whereas the crystallographically inequivalent 4a site remains magnetically silent in this material.
\end{abstract}

\pacs{75.30.Mb 71.27.+a 78.70.Nx\vspace{-3pt}}

\maketitle\enlargethispage{4pt}

Magnetic dynamics in heavy-fermion metals usually represent an intricate tangle of the local-moment fluctuations and the spin-dynamical response of itinerant heavy quasiparticles \cite{Stewart84, FuldeLoewenhaupt86}. The strong hybridization of the localized 4$f$ electron states with the conduction band makes these two contributions difficult to decouple. For instance, in the most classical heavy-fermion compounds like CeCu$_6$, CeRu$_2$Si$_2$ and CeAl$_3$, the low-temperature inelastic neutron scattering (INS) signal consists of a momentum-independent single-site quasielastic magnetic scattering (QEMS) attributed to localized Kondo-type excitations and an inelastic contribution from inter-site magnetic correlations due to Ruderman-Kittel-Kasuya-Yosida (RKKY) interactions, which merge together at higher temperatures \cite{AeppliYoshizawa86, RossatMignod88, LazukovAlekseev02+TidenAlekseev07}.

Despite these qualitative similarities, the momentum-space distribution of the QEMS response in the paramagnetic state is strongly material dependent. Thus, CeRu$_2$Si$_2$ exhibits incommensurate magnetic fluctuations peaked at $(0.3~0~0)$ and $(0.3~0.3~0)$ wave vectors \cite{RossatMignod88}, while short-range antiferromagnetic (AFM) correlations were found near $(1\,0\,0)$ in CeCu$_6$ \cite{AeppliYoshizawa86, RossatMignod88}. The resulting momentum-space structure of the zero-frequency susceptibility, $\chi_0(\mathbf{Q},\omega\!=\!0)$, carries essential information about the material's electronic properties and its tendency to magnetic instabilities driven by the RKKY coupling. From more recent examples, a direct relationship between the Fermi-surface nesting properties and the short-range magnetic correlations, resulting in a diffuse neutron-scattering signal, was demonstrated for Tb$_2$PdSi$_3$~\cite{InosovEvtushinsky09}. Of particular relevance for the present work are also our recent results on CeB$_6$ \cite{FriemelLi12, JangFriemel14}, where a maximum in the normal-state QEMS intensity was found to coincide in momentum space with the $(\half\half\half)$ propagation vector of the magnetically hidden order that sets in below $T_\text{Q}=3.2$\,K and is usually attributed to the antiferroquadrupolar (AFQ) ordering of the localized Ce\,4$f$ quadrupolar moments \cite{EffantinRossatMignod85, SantiniCarretta09+KuramotoKusunose09}. At even lower temperatures below $T_\text{N}=2.3$\,K, it succumbs to a multi-$\mathbf{k}$ commensurate AFM order \cite{EffantinRossatMignod85, ZaharkoFischer03}, which stabilizes a narrow band of dispersive magnetic excitations in the INS response \cite{FriemelLi12, JangFriemel14} that can be explained by the formation of a low-energy resonant spin-exciton mode \cite{AkbariThalmeier12}.

The magnetic phase diagram of Ce$_3$Pd$_{20}$Si$_6$ \cite{MitamuraTayama10, OnoNakano13, PaschenLarrea14} nearly replicates that of CeB$_6$, though with reduced characteristic temperature and magnetic-field scales. The AFM ordering temperature does not exceed $T_\text{N}=0.31$\,K in the highest-quality stoichiometric samples, according to a corpus of available studies performed on both single crystals \cite{ProkofievCusters09, MitamuraTayama10, OnoNakano13} and powders \cite{TakedaKitagawa95, DuginovGritsaj00, StrydomPikul06, PaschenMueller07, GotoWatanabe09, CustersLorenzer12}. The AFM phase can also be suppressed by a magnetic field of only 0.7\,T applied along the $[100]$ crystallographic direction \cite{MitamuraTayama10, OnoNakano13}. This places Ce$_3$Pd$_{20}$Si$_6$ very close to a quantum critical point (QCP) \cite{StrydomPikul06, PaschenMueller07}, which has been reached in polycrystals by the application of a magnetic field of less than 1~T \cite{CustersLorenzer12}, and which is likely accessible also under a small hydrostatic or chemical pressure \cite{ProkofievCusters09}. Such a proximity leads to non-Fermi-liquid behavior and, in particular, to very high values of the electronic specific-heat coefficient, $\gamma=\lim_{T\rightarrow0}{\scriptsize\Delta}C(T)/T$. Reportedly, this can reach up to $\sim$\,8\,J/(mol$_\text{Ce}\cdot$K$^2$) near the QCP, making Ce$_3$Pd$_{20}$Si$_6$ one of the heaviest-electron systems known to date \cite{TakedaKitagawa95, CustersLorenzer12}.

Like in CeB$_6$, the AFM phase in Ce$_3$Pd$_{20}$Si$_6$ is surrounded by the so-called phase~II, which is also attributed to an AFQ order \cite{MitamuraTayama10, OnoNakano13} yet remains much less studied. In zero field, this phase persists only in a narrow temperature window between $T_\text{N}$ and \mbox{$T_\text{Q}\approx0.45$\,--\,0.5\,K}, according to thermodynamic measurements \cite{StrydomPikul06, MitamuraTayama10, OnoNakano13}. It is initially stabilized by the application of small magnetic fields of a few teslas, but is eventually suppressed at even higher fields, leading to another qualitative similarity to the AFQ phase of CeB$_6$ \cite{GoodrichYoung04}, albeit in a much more accessible field range.

\begin{figure*}[t]\vspace{-1em}
\includegraphics[width=\textwidth]{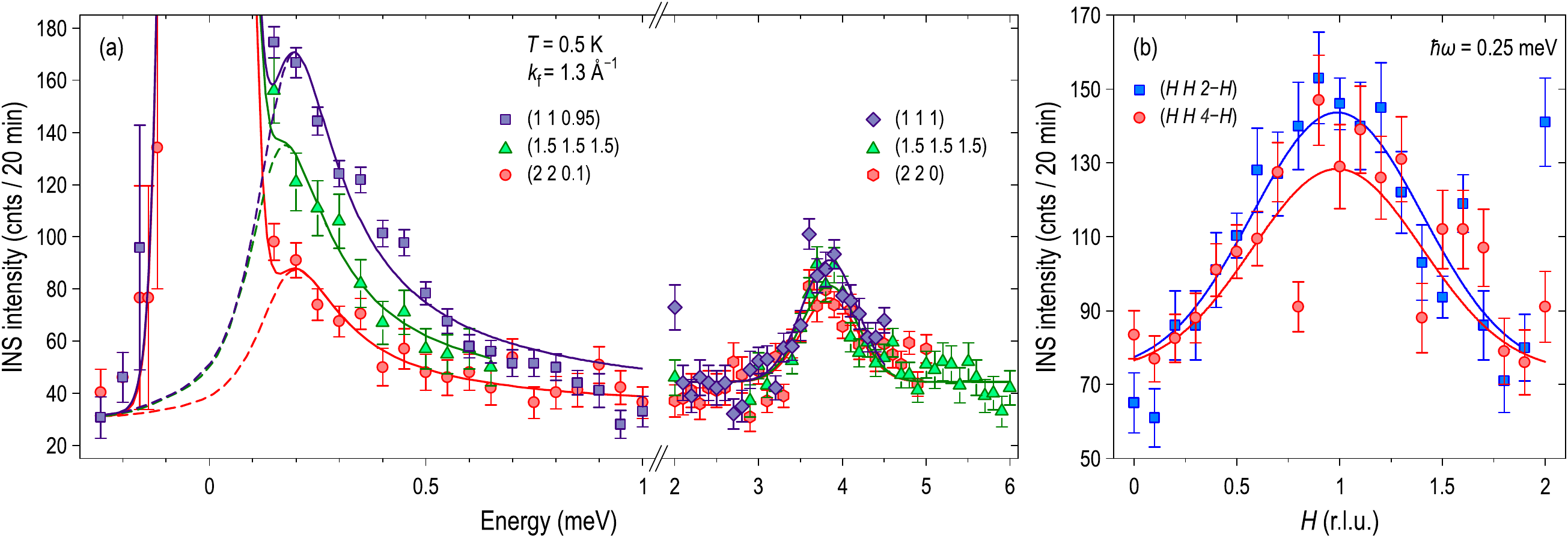}
\caption{(a)~Energy dependence of the magnetic scattering at base temperature ($T\approx0.5$\,K), measured at several wave vectors, as indicated in the legend. The left side of the panel shows the $\mathbf{Q}$-dependent QEMS response, fitted to the quasielastic Lorentzian line shape as given by Eq.\,(\ref{Eq:Quasielastic}), whereas the right-hand-side presents data measured across the $\mathbf{Q}$-independent CEF line centered at 3.9\,meV, fitted to Gaussian profiles. The dashed lines show the magnetic signal without the elastic incoherent scattering contribution. (b)~Momentum dependence of the signal at $\hslash\omega=0.25$\,meV along two equivalent trajectories: ($H$\,$H$\,2\,--\,$H$) and ($H$\,$H$\,4\,--\,$H$), fitted with Gaussian profiles.}
\label{Fig:Qdep}\vspace{-2pt}
\end{figure*}

The very low temperature scales and complications due to the intricate crystal structure of Ce$_3$Pd$_{20}$Si$_6$ have so far precluded any direct observations of the AFQ order by diffraction methods, either with or without the application of magnetic field. As a result, the propagation vector of phase~II remains unknown, whereas in the structurally much simpler CeB$_6$ it has been determined as $\mathbf{q}_{\rm AFQ}=(\half\half\half)$ both by neutron scattering \cite{EffantinRossatMignod85, FriemelLi12} and resonant x-ray diffraction \cite{CeB6resonantXray}. One essential complication that hinders similar measurements on Ce$_3$Pd$_{20}$Si$_6$ is that its crystal structure includes two interpenetrating sublattices of Ce ions on crystallographically inequivalent 4a (Ce1) and 8c (Ce2) Wyckoff sites in a cubic unit cell with the $Fm\bar{3}m$ space group (for an illustration, see e.g.~Ref.~\citenum{CustersLorenzer12}). This results in the unit cell parameter of the simple cubic Ce2 sublattice being half that of the face centered cubic Ce1 sublattice, $a=12.28$\,\AA. As a consequence, additional magnetic Bragg reflections due to an AFQ order of the same kind as in CeB$_6$, residing on the Ce2 sublattice, would coincide with the much stronger $(111)$ structural Bragg reflections and be therefore much more difficult, if not impossible, to observe directly.

Previous INS measurements on Ce$_3$Pd$_{20}$Si$_6$ have been performed, to the best of our knowledge, only on polycrystalline samples \cite{PaschenLaumann08, DeenStrydom10}. They revealed a clear crystalline electric field (CEF) line at 3.9\,meV \cite{PaschenLaumann08} and suggested the presence of an additional unresolved low-energy peak centered at 0.31\,meV, which reportedly persisted up to ambient temperature and was attributed to another CEF excitation \cite{DeenStrydom10}. However, follow-up measurements performed on the same powder sample with a better energy resolution, which we will present further on, show a clear magnetic signal centered at much lower energies, consistent with a quasielastic response.

In the present study, we investigated the low-energy spin dynamics of Ce$_3$Pd$_{20}$Si$_6$ using single-crystal INS spectroscopy. For triple-axis (TAS) measurements, we used one large single crystal with a mass of 1.89~g, whereas for the time-of-flight (TOF) experiment, it was coaligned with an additional larger crystal, resulting in the total sample mass of $\sim$\,5.9\,g. Both crystals were characterized by resistivity measurements, indicating a sharp magnetic transition at $T_{\rm N}=0.23$\,K. The crystals were mounted on a copper sample holder in the $(HHL)$ scattering plane to allow access to all high-symmetry directions of the cubic Brillouin zone (BZ). The mosaic spread of the sample, determined from the full width at half maximum (FWHM) of the rocking curves measured on structural Bragg reflections during sample alignment, was better than 0.5$^\circ\!$. For the low-temperature TAS measurements, the sample was first mounted in a $^3$He/$^4$He dilution refrigerator \cite{DilutionFridge}. Afterwards, for $T$-dependent measurements, the sample was remounted in a conventional $^4$He closed-cycle refrigerator with an exchange gas. The TAS measurements were taken using the 4F2 cold-neutron spectrometer at the Laboratoire L\'eon Brillouin (LLB), Saclay, France, operated with the fixed final neutron wave vector $k_\text{f}=1.3$\,\AA$^{-1}$ that corresponds to an energy resolution of 0.11\,meV, defined as the FWHM of the elastic line.

The unprocessed low-temperature energy scans at several representative wave vectors are shown in Fig.\,\ref{Fig:Qdep}\,(a). At low energies, below 1\,meV, we observe a QEMS signal that can be described by a quasielastic Lorentzian line shape \cite{RobinsonRadousky_book},\vspace{-2pt}
\begin{multline}\label{Eq:Quasielastic}
S(\mathbf{Q},\omega)\propto F^2(\mathbf{Q})\,\frac{\chi_0(\mathbf{Q})}{1-\exp(-\hslash\omega/k_\text{B}T)}\\
\times\frac{\omega}{2\piup}\biggl(\frac{\Gamma}{\hslash^2(\omega-\omega_0)^2+\Gamma^2}+\frac{\Gamma}{\hslash^2(\omega+\omega_0)^2+\Gamma^2}\biggr).
\end{multline}
Here $F(\mathbf{Q})$ is the Ce$^{3+}$ magnetic form factor, $\chi_0(\mathbf{Q})$ is the momentum-dependent static susceptibility, and $\Gamma$ is the half-width of the Lorentzians centered at $\pm\hslash\omega_0$. This signal has nearly identical shape in energy at different wave vectors, but its intensity varies strongly in $\mathbf{Q}$-space, as evidenced by a twofold difference in the Lorentzian amplitude between the $(111)$ and $(220)$ points. Note that the measurements were done at slightly incommensurate wave vectors to avoid the contamination from phonons and the Bragg tail.

At higher energies, one can see a CEF line centered at 3.9\,meV, already known from previous INS measurements on powder samples \cite{PaschenLaumann08, DeenStrydom10}. As expected, it exhibits no dispersion and its intensity is nearly constant in momentum space apart from a minor form-factor suppression towards higher $|\mathbf{Q}|$. This qualitatively different behavior distinguishes it from the QEMS signal, characterized by a strongly $\mathbf{Q}$-dependent dynamical structure factor. At the same time, the low-energy CEF line at $\sim$\,0.31~meV suggested in Ref.\,\citenum{DeenStrydom10} from data with a lower energy resolution can be clearly excluded by our present measurements.

\begin{figure}[t]
\includegraphics[width=\columnwidth]{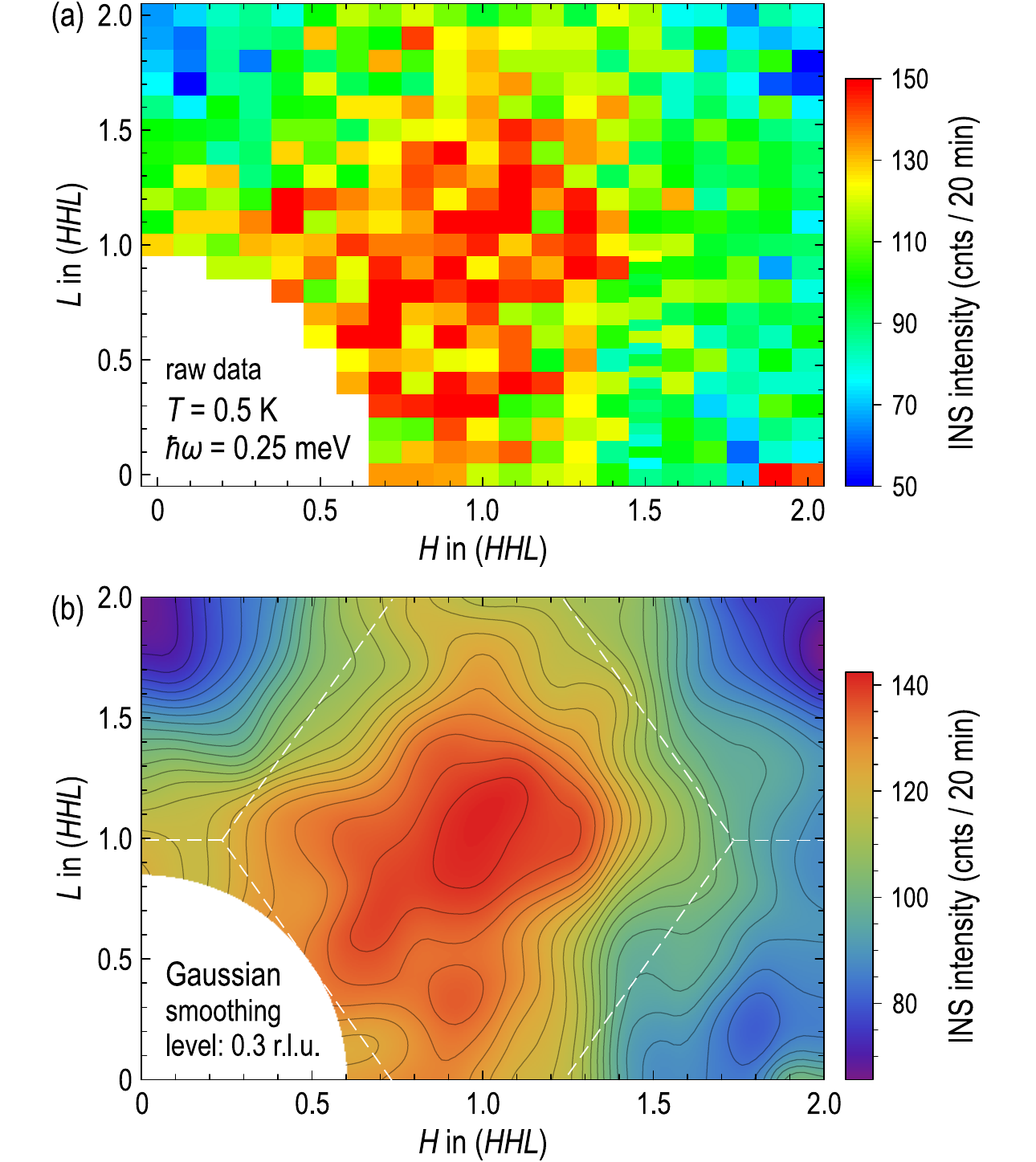}
\caption{Momentum dependence of the low-$T$ QEMS intensity in the $(HHL)$ scattering plane, measured at an energy transfer of 0.25\,meV: (a)~raw INS data; (b)~a contour map showing the same data smoothed with a two-dimensional Gaussian filter characterized by the FWHM of 0.3\,r.l.u. White dashed lines mark BZ boundaries of the face-centered cubic Ce$_3$Pd$_{20}$Si$_6$ lattice.\vspace{-1pt}}
\label{Fig:ColorMaps}\vspace{-2pt}
\end{figure}

Two constant-energy scans measured at 0.25\,meV along equivalent Brillouin-zone diagonals, as presented in Fig.\,\ref{Fig:Qdep}\,(b), demonstrate a broad $\mathbf{Q}$-space distribution of the quasielastic intensity, peaked at the $(111)$ wave vector. From here on we show momentum in reciprocal lattice units ($1\,\text{r.l.u.}~= 2\piup/a$ with $a=12.28$\,\AA). The peak width of $\,\sim\kern.5pt$1\,r.l.u. is suggestive of short-range dynamical AFM correlations over distances of the order of one lattice constant, or two interatomic distances of the Ce\,8c sublattice. The momentum scans along ($H$\,$H$\,2\,--\,$H$) and ($H$\,$H$\,4\,--\,$H$) are essentially identical apart from a small form-factor suppression of intensity at higher $|\mathbf{Q}|$, which confirms that the periodicity of the signal in momentum space matches with that of the Ce\,8c sublattice, i.e. a translation by a vector with all even Miller indices results in an equivalent $\mathbf{Q}$ vector. On the other hand, the $(111)$ and $(002)$ points that are expected to be equivalent for the Ce~4a sublattice show different intensity, indicating that the QEMS response breaks the symmetry of the face-centered cubic BZ and should therefore originate predominantly from magnetic correlations on the Ce\,8c sites. This situation is reminiscent of that in iron pnictides, where spin fluctuations inherit the symmetry of the unfolded BZ because of the higher symmetry of the magnetic Fe sublattice with respect to the crystal itself \cite{ParkInosov10}.

\begin{figure}[b]
\includegraphics[width=\columnwidth]{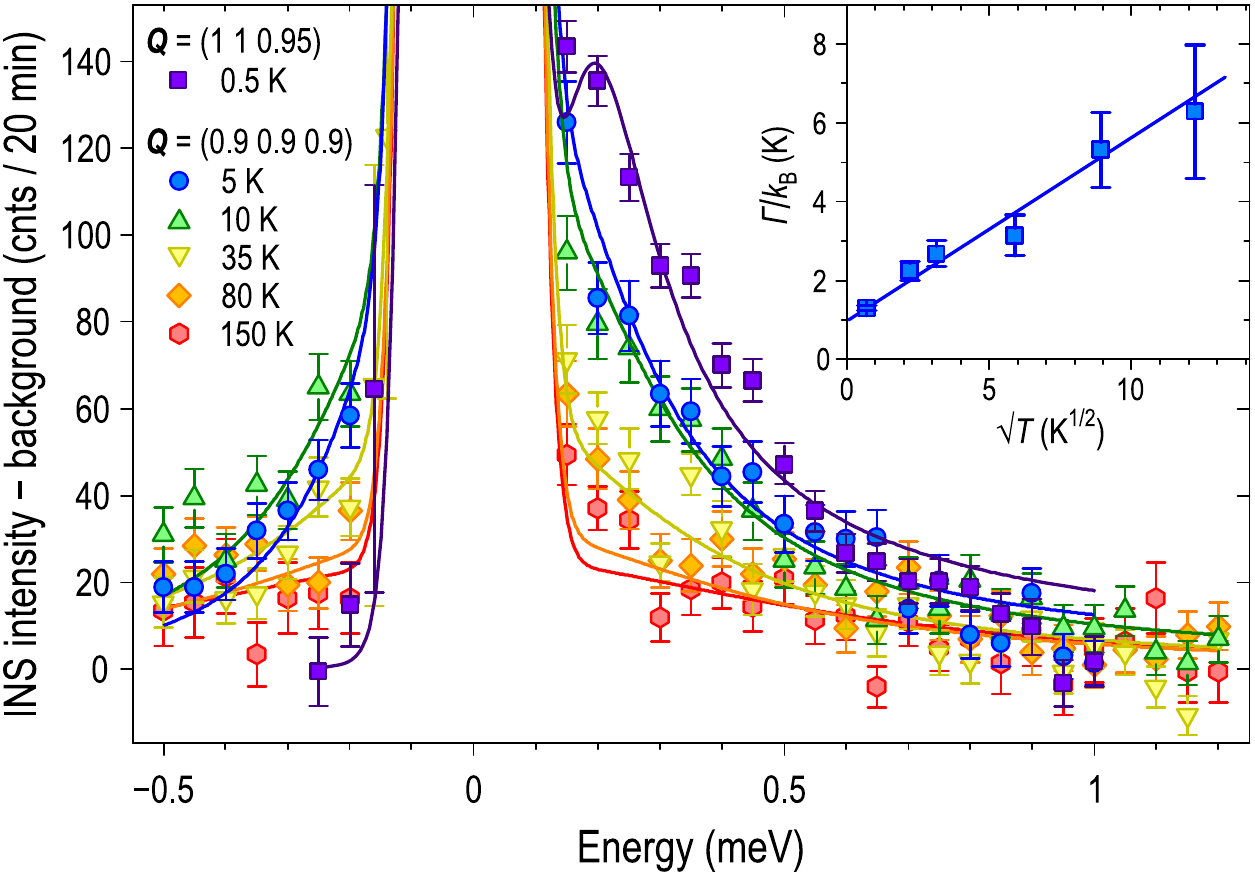}
\caption{Temperature evolution of the background-subtracted QEMS intensity, fitted to the sum of a quasielastic Lorentzian (Eq.\,\ref{Eq:Quasielastic}) and the incoherent elastic line. The inset shows the $T$-dependence of the normalized quasielastic line width, $\Gamma(T)/k_\text{B}$, plotted vs. $T^{1/2}$ to emphasize the $\Gamma_0+A\sqrt{T}$ dependence.}
\label{Fig:Tdep}\vspace{-4pt}
\end{figure}

A more complete picture of the quasielastic intensity distribution in $\mathbf{Q}$-space is given by Fig.\,\ref{Fig:ColorMaps}, showing a constant-energy map of the low-temperature QEMS response at 0.25\,meV over the entire $(HHL)$ scattering plane. From the fact that the quasielastic line shape remains essentially unchanged with $\mathbf{Q}$, as follows from Fig.\,\ref{Fig:Qdep}, we can conclude that such an intensity map is also representative of the total energy-integrated spectral weight distribution in momentum space. It shows a broad anisotropic hump of intensity centered at $(111)$, with weaker side lobes extending along the $(110)$ and $(001)$ directions. In the large BZ corresponding to the Ce\,8c sublattice, this wave vector would coincide with the zone corner ($R$ point), matching with the AFQ propagation vector of CeB$_6$, where a similarly broad local maximum of the QEMS intensity was also found above $T_\text{N}$ \cite{FriemelLi12}. Yet, the lowest intensity in our dataset is observed in the vicinity of the $(002)$, $(220)$ and $(222)$ wave vectors, i.e. near the center of the large BZ. This is in remarkable contrast to CeB$_6$, which hosts strong ferromagnetic fluctuations at these points \cite{JangFriemel14}.

Next, we consider the temperature dependence of the QEMS response near its maximum at the $(111)$ wave vector, as shown in Fig.\,\ref{Fig:Tdep}. Here, all but the lowest-temperature datasets were measured in the closed-cycle $^4$He refrigerator. As this cryostat produced a different background from the one observed with the $^3$He/$^4$He dilution fridge, the data in Fig.\,\ref{Fig:Tdep} are plotted after subtraction of the corresponding constant background levels (shared for all data measured under the same conditions). Upon warming, we can observe a monotonic suppression and broadening of the quasielastic signal. The temperature dependence of the quasielastic line width, $\Gamma(T)$, presented in the inset to Fig.\,\ref{Fig:Tdep}, follows the conventional $T^{1/2}$ law \cite{RobinsonRadousky_book},\vspace{-3pt}\enlargethispage{3pt}
\begin{equation}
\Gamma(T)/k_\text{B}=\Gamma_0/k_\text{B}+A\kern.5pt\sqrt{T}.\vspace{-2pt}
\end{equation}
From the residual width at \mbox{$T=0$}, the characteristic neutron-deduced Kondo temperature, \mbox{$T_\text{K}=\Gamma_0/k_\text{B}=(0.97\pm0.07)$\,K}, can be inferred.


\begin{figure}[t]
\includegraphics[width=\columnwidth]{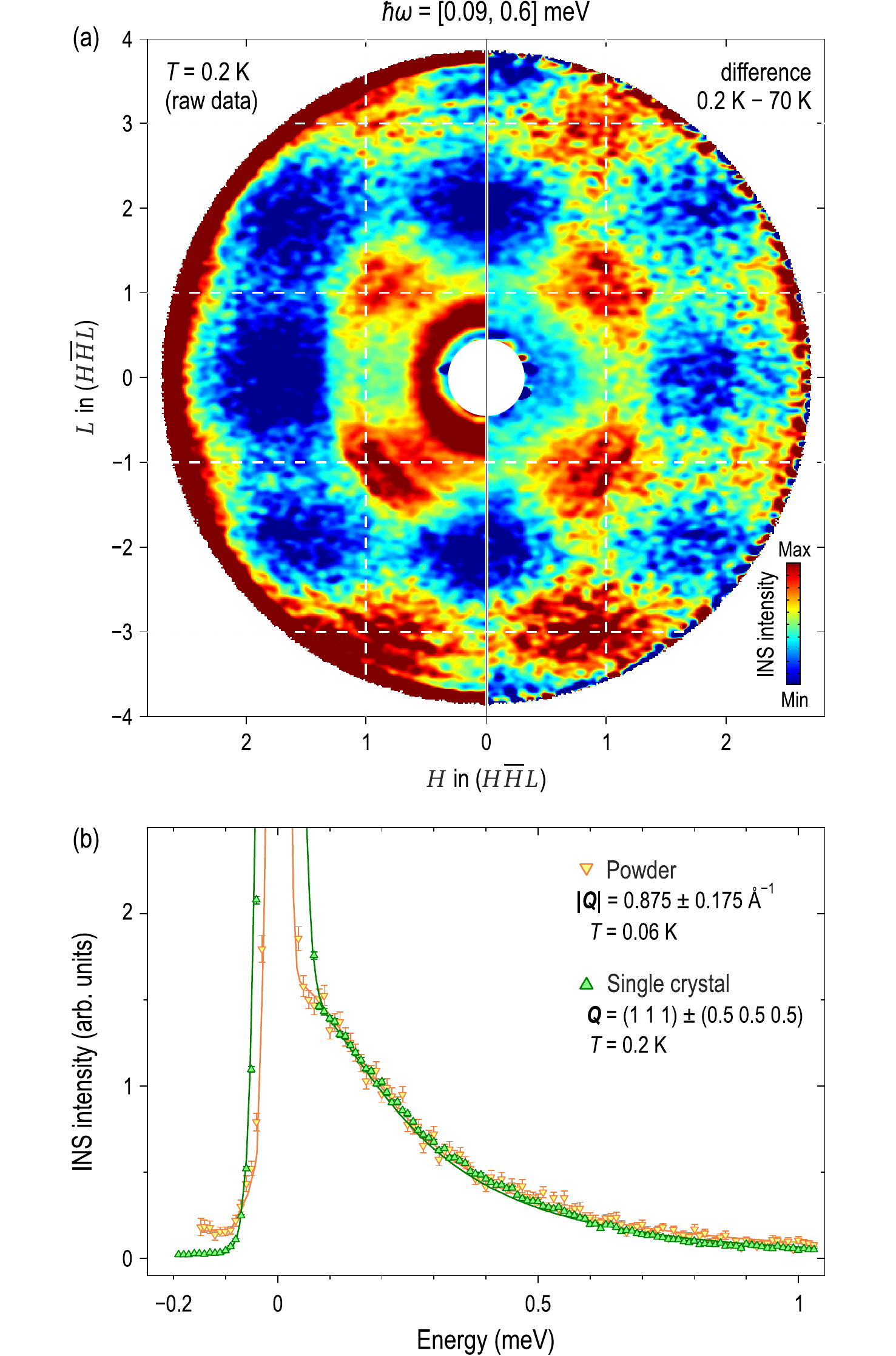}
\caption{(a)~Momentum dependence of the QEMS intensity in the $(HHL)$ scattering plane, obtained by integrating the TOF data in the energy window $0.09\,\text{meV} \leq \hslash\omega \leq 0.6\,\text{meV}$. The left panel shows raw data collected at $T\approx0.2$\,K, whereas the right panel presents the same dataset after subtraction of the high-temperature nonmagnetic background measured at $T=70$\,K. White dashed lines mark BZ boundaries corresponding to the simple cubic Ce\,8c sublattice. (b)~High-resolution energy profiles of the QEMS intensity from powder and single-crystal samples, integrated around the $(111)$ wave vector, as indicated in the legend.}
\label{Fig:TOF-Edep}\vspace{-2pt}
\end{figure}

To get a broader overview of the $\mathbf{Q}$-space and to ensure that no additional magnetic contributions are present at lower energy transfers, we also performed TOF measurements on a larger sample using the cold-neutron chopper spectrometer (CNCS) \cite{EhlersPodlesnyak11} at the Spallation Neutron Source (Oak Ridge National Laboratory, USA). The energy of the incident neutrons was set to $E_{\rm i}=2.49$\,meV, providing an energy resolution with a FWHM of 0.051\,meV, that is twice better than in the TAS experiment. The dataset collected at $T=0.2$\,K (setpoint value) was complemented by a higher-temperature background measurement at 70\,K, where the low-energy magnetic spectral weight is considerably reduced, according to Fig.\,\ref{Fig:Tdep}. The corresponding intensity maps of the $(HHL)$ plane before and after subtraction of the high-temperature background, integrated immediately above the elastic line in the energy range $0.09\,\text{meV} \leq \hslash\omega \leq 0.6\,\text{meV}$, are presented in Fig.\,\ref{Fig:TOF-Edep}\,(a). In agreement with Fig.\,\ref{Fig:ColorMaps}, we see broad intensity maxima at wave vectors with all odd Miller indices, i.e. at the corners of the large BZ corresponding to the Ce\,8c sublattice. Down to the lowest accessible energy (0.09\,meV), we find neither any additional magnetic contributions that could be reconciled by symmetry with the face-centered Ce\,4a sublattice, nor any ferromagnetic fluctuations like those found in CeB$_6$ at the zone center \cite{JangFriemel14}.\enlargethispage{3pt}

In Fig.\,\ref{Fig:TOF-Edep}\,(b), we also compare energy profiles of the QEMS signal, obtained from our TOF data by integration within $\pm0.5$~r.l.u.\hspace{-0.6pt} on either side of the $(111)$ wave vector, to the corresponding energy dependence measured earlier on a powder sample \cite{DeenStrydom10} at the IRIS spectrometer (ISIS, UK) with an even better energy resolution of 0.025\,meV. The perfect agreement between the two curves confirms that the signal is sample-independent and only originates from the $(111)$ fluctuations. The exact form of the signal deviates from the perfect Lorentzian line shape, but could be reconciled with a generalized Voigt profile, shown as solid lines.

To summarize, our results demonstrated the presence of low-energy dynamical magnetic correlations in the paramagnetic state of Ce$_3$Pd$_{20}$Si$_6$, which could be responsible for the excess magnetic entropy in specific heat \cite{TakedaKitagawa95}. According to their $\mathbf{Q}$-space symmetry, they are associated with the same simple-cubic Ce\,8c sublattice that was shown earlier to host static AFM order below $T_{\rm N}$ \cite{LorenzerStrydom14}. This suggests that the remaining Ce\,4a ions are magnetically inactive, which could be either due to the frustration on the face-centered cubic sublattice, strong Kondo screening of their magnetic moments, or both. The possibly large difference in the Kondo scales on different sublattices would be in line with the theoretically suggested regime of competing Kondo effects \cite{BenlagraFritz11}. Despite the strikingly similar magnetic phase diagrams of Ce$_3$Pd$_{20}$Si$_6$ and CeB$_6$, both exhibiting an AFQ phase, their spin-fluctuation spectra are markedly different: Ferromagnetic correlations that dominate the spectrum of CeB$_6$ are absent in Ce$_3$Pd$_{20}$Si$_6$, while the dynamical AFM correlations in Ce$_3$Pd$_{20}$Si$_6$ are much more short-range and extend over distances of only about one lattice constant. Nevertheless, from the presence of strong quasielastic scattering at the BZ corner in both compounds, which coincides in CeB$_6$ with the propagation vector of the AFQ phase, we may tentatively surmise that the AFQ phase in Ce$_3$Pd$_{20}$Si$_6$ may also reside at the same wave vector in the large BZ, which is $\mathbf{q}_{\rm AFQ}=(111)$.

\emph{Acknowledgments.} We are grateful to Ph. Boutrouille (LLB) and S. Elorfi (SNS) for technical support during the experiments. Reduction of the TOF data was done using the \emph{Horace} software package \cite{Horace}. This project was funded by the German Research Foundation (DFG) under grant No.~IN\,209/3-1 and via the Research Training Group GRK\,1621 at the TU Dresden, by the European Research Council (Advanced Grant QuantumPuzzle, No. 227378), and by the European Commission under the 7$^\text{th}$ Framework Programme NMI3-II/FP7\,---\,Contract No. 283883. Research at Oak Ridge National Laboratory's Spallation Neutron Source was supported by the Scientific User Facilities Division, Office of Basic Energy Sciences, US Department of Energy.


\begin{thebibliography}{99}
\bibitem{DilutionFridge} In this configuration, the sample was fixed to a thermometer block with no temperature sensor attached directly to the sample. Consequently, due to the absence of an exchange gas, the real sample temperature could be somewhat higher than the setpoint value. From the absence of magnetic Bragg scattering \cite{LorenzerStrydom14} we conclude that the actual sample temperature was above $T_{\rm N}$, hence the base-temperature data are representative of the paramagnetic state. We additionally confirmed this by checking that no detectable change in the inelastic scattering intensity could be seen upon increasing the temperature to 0.9\,K~$\gg\!T_{\rm N}$.
\end{thebibliography}
\end{document}